\documentclass{ieee}
\usepackage{times}
\usepackage{graphicx}
\usepackage{amssymb,amsmath,amsthm}
\theoremstyle{plain}
  \newtheorem{theorem}{Theorem}
  \newtheorem{lemma}[theorem]{Lemma}
  
\theoremstyle{definition}
  \newtheorem{definition}[theorem]{Definition}

\theoremstyle{remark}

\begin{document}
  
  \title{Estimating seed sensitivity on homogeneous alignments}
  
  \author{
    Gregory Kucherov\\
    LORIA/INRIA-Lorraine,\\
    615, rue du Jardin Botanique,\\ 
    B.P. 101, 54602\\
    Villers-l\`es-Nancy France,\\ 
    Gregory.Kucherov@loria.fr\\
    \and
    Laurent No{\'e}\\
    LORIA/INRIA-Lorraine,\\
    615, rue du Jardin Botanique,\\
    B.P. 101, 54602\\
    Villers-l\`es-Nancy France,\\
    Laurent.Noe@loria.fr\\
  \and
  Yann Ponty\\
  LRI, UMR CNRS 8623,\\
  B\^at 490 Universit\'e Paris-Sud\\
  91405 Orsay, France,\\
  Yann.Ponty@lri.fr
  }

\maketitle
\thispagestyle{empty}

\begin{abstract}
  We address the problem of estimating the sensitivity of seed-based
  similarity search algorithms. In contrast to 
  approaches based on Markov models
  \cite{PATTERNHUNTER02,BuhlerKeichSunRECOMB03,BrejovaBrownVinarCPM03,BrejovaBrownVinarWABI03,ChoiZhang03},
  we study the estimation based on {\em homogeneous} alignments. We
  describe an algorithm for counting and random generation of those
  alignments and an algorithm for exact computation of the sensitivity
  for a broad class of seed strategies. We provide experimental
  results demonstrating a bias introduced by ignoring the homogeneousness
  condition.
\end{abstract}
\section{Introduction}
Comparing nucleic acid or protein sequences remains by far the most
common bioinformatics application. The classical local alignment
problem consists in computing most significant similarities between
two sequences, or between a sequence and a database. 
The significance of an alignment is measured by a {\em score},
commonly defined using
an additive principle by assigning a 
positive score to each matching character, and a negative
score (penalty) to each mismatch and each contiguous gap. 

Best-scoring local alignments can be computed by the well-known
Smith-Waterman dynamic programming algorithm \cite{SW81}, however for
large-scale sequence comparison this computation becomes too
time-consuming. Several heuristic algorithms have been designed to
speed up the computation of local alignments, at the price of
possibly missing some alignments or computing their sub-optimal
variants. BLAST \cite{BLAST90} is the most prominent representative of this
family, Fasta \cite{FASTA88} is another example. Both these programs are
based on the common principle: similarity regions are assumed to
share one or several short fragments, called {\em seeds}, 
that are used to detect potential similarities.
More recently, it has been understood that
using non-contiguous (called also {\em spaced} or {\em gapped}) seeds instead of
contiguous substrings can considerably improve the
sensitivity/selectivity trade-off. 
PatternHunter~\cite{PATTERNHUNTER02} was the first method that used carefully
designed spaced seeds to improve the sensitivity of DNA local
alignment. Spaced seeds have been also shown to improve the efficiency
of lossless filtration for approximate pattern
matching~\cite{PevznerWaterman95,BurkhardtKarkkainen03}. Earlier, random spaced
seeds were used in FLASH program \cite{FLASH93} to cover
sequence similarities, and the sensitivity of this approach was recently studied in
\cite{LSHALLPAIRS01}. For the last two years, spaced seeds received an increasing
interest \cite{BuhlerKeichSunRECOMB03,BrejovaBrownVinarCPM03,BrejovaBrownVinarWABI03,ChoiZhang03,CZZ03} and
have been used in new local alignment tools
\cite{BLASTZ03,NoeKucherovRRINRIA03}. 



Coming back to the concept of alignment score, note that all
heuristic algorithms typically try to output alignments with a
score greater than some user-defined bound. However, computing 
{\em all} such alignments would be an
ill-defined task. Usually, the alignments of interest are those
which, on the one hand, 
do not
contain in them big regions of
negative score (in which case the alignment should probably be split
into two or more alignments of higher score) and on the other hand,
are not too short to be a part of a larger high-scoring similarity. This
is captured by the Xdrop heuristics, a part of BLAST
algorithm: once a seed has been  identified, the Xdrop algorithm
extends it in both directions into so-called High-scoring Segment Pair (HSP),
as long as the ``running score'' 
does not decrease more than by a certain value. All other seed-based algorithms
apply a similar approach in that they extend the found seed (or group
of seeds) outside as far as the total score does not undergo a
prohibitive drop. 

The alignments found in such a way are formalized through the notion
of {\em maximal scoring segment}\footnote{We use the term 
{\em segment} instead of {\em subsequence} \cite{RuzzoTompaISMB99} as the
latter usually does not require the elements to be contiguous.}
\cite{RuzzoTompaISMB99}. Consider a {\em gapless alignment} which can be
naturally translated into a sequence of match/mismatch scores. We call
this alignment (or sequence) {\em homogeneous} if it does not contain
a proper contiguous subalignment (segment) whose score is greater than
that of the whole alignment. 
Given an alignment (or sequence), a homogeneous subalignment is
called a {\em maximal scoring subalignment (segment)} 
if it is extended to the right and left as
far as the homogeneity property is verified. In other words, a maximal
scoring segment is not included in a larger homogeneous segment. 

Abstracting from possible gaps, alignments found by similarity search
programs are maximal scoring segments. On the other hand,
maximal scoring segments capture a biologically relevant notion of
alignment: if an alignment is not homogeneous then it is likely to be a
merge of two alignments that should be considered as distinct, and if
an alignment is not maximal, it is likely to be a 
part of a larger interesting alignment. 

In this context, the main motivation of this work can be summarized by
the following claim: {\em Since homogeneous alignments are those which are
  really found and intended to be found by similarity search
  algorithms, then the efficiency of those algorithms has to
  be measured on homogeneous alignments rather than on arbitrary
  alignments.} 

With this motivation in mind, we propose an approach to 
analyze the sensitivity
of similarity search algorithms. Using this approach, we
demonstrate that measuring the sensitivity of algorithms on general
alignments instead of homogeneous alignments (as it is usually done)
leads to biased results, more specifically to an underestimation of
the sensitivity. 

In this paper,
we propose a dynamic programming algorithm to compute the
sensitivity estimator (hit probability) with respect to homogeneous
alignments. The algorithm, which is an extension of algorithms
proposed in
\cite{KeichLiMaTromp02,BuhlerKeichSunRECOMB03,BrejovaBrownVinarCPM03},
works for a wide range of seed definitions 
(contiguous or spaced seeds, single- double- or multiple-seed
approaches, 
etc) and therefore can guide
the choice of the seed strategy. It is based on the
enumeration of homogeneous alignments. On the other hand, the
enumeration allows us to obtain an efficient random generation
algorithm for homogeneous alignments. The latter, in turn, can be
used for estimating the hit probability experimentally by testing
the chosen seed criterion on a large number of random homogeneous
alignments. 

Finally, note that the homogeneous sequences have
other applications in biological sequence analysis. Karlin, Altschul 
and other authors \cite{KarlinBrendelSCIENCE92,KarlinAltschulPNAS93} studied long homogeneous (and maximal
scoring) segments in protein sequences, where each residue is assigned
a score according to a certain scoring function. It has been
demonstrated that those segments are often biologically significant,
and may, for example, correspond to transmembrane domains
\cite{KarlinAltschulPNAS93}. Therefore, the methods proposed in our paper can potentially 
apply to other bioinformatics problems than the problem of measuring
the sensitivity of local alignments programs considered here. 

The paper is organized as follows. Section~\ref{enumeration} 
presents algorithms, based on enumeration techniques, for the uniform
random generation of homogeneous sequences. 
Section~\ref{sensitivity} describes an exact algorithm to compute the
seed detection probability on homogeneous
sequences. Section~\ref{experiments} is devoted to experiments, and
demonstrates the bias induced by considering general sequences rather than
homogeneous ones. Finally, Section~\ref{conclusions} discusses
possible extensions and directions for future work.

\section{Enumeration and random generation of homogeneous sequences}
\label{enumeration}

Our main object of study is the {\em gapless alignment} of two DNA
sequences. We represent it by a binary sequence
$A=(b_1,\ldots,b_n)$, $b_i\in\{0,1\}$, where $1$ stands for a match
and $0$ for a mismatch. 
We assume that each match is assigned
a constant positive integer score $s$, and each mismatch a constant negative
integer score (penalty) $-p$, regardless of the mismatching letters. 
This is the case for many
nucleotide scoring systems, for example a popular BLAST default
scoring system assigns 1 to each match and -3 to each
mismatch.\footnote{More accurate scoring systems assign different
  penalties to different mismatches. In particular, it is very useful
  to distinguish between transitions (substitutions
  $\mathtt{A}\leftrightarrow\mathtt{G}$ and 
  $\mathtt{C}\leftrightarrow\mathtt{T}$) and transversions (other
  substitutions), since transition mutations occur with a greater
  relative frequency than transversions, particularly in coding
  sequences. This distinction is also useful in seed design,
as it allows to consider transition-constrained
  mismatches, as done by BLASTZ \cite{BLASTZ03} or YASS \cite{NoeKucherovRRINRIA03} for example. A further
  step would be to make finer partitions of all nucleotide pairs, depending
  on statistical properties of analyzed sequences. In our setting,
  this would imply the modeling of alignments by
  sequences over three or more letters. We will discuss this extension
  in Section~\ref{conclusions}.
  }\label{page1}
An alternative representation of $A$ is then the sequence of
individual scores, called {\em score sequence}, $X_A=(x_1,\ldots,x_n)$,
$x_i\in \{s,-p\}$ and $x_i=-p+b_i(s+p)$, and the 
score of the whole alignment is $S(X)=\sum_{k=1}^n x_k$. 
\begin{definition}
A binary sequence $A$ (and the associated score sequence $X_A$)
is called {\em homogeneous}  
if $S(X_A)$ is strictly greater than $S(X_A[i..j])$ for all proper segments
$X_A[i..j]=(x_i,\ldots,x_j)$ ($i>1$ or $j<n$). 
\end{definition}

This section is devoted to the following question: How to uniformly
generate random homogeneous sequences of a given length $n$, or 
of given length $n$ and score $S$? Here the uniformity
condition means that each sequence of the considered class has the
same probability to occur.

The interest of this question is twofold. First, being able to
randomly generate homogeneous sequences would allow to measure the
sensitivity of a similarity search algorithm in the experimental way, 
by testing it against a sample of randomly drawn homogeneous
sequences. Second, as it is often the case for combinatorial objects,
the question of random generation of homogeneous sequences is
closely related to the question of their enumeration, and the
latter will be used in the next section to obtain an exact algorithm
for computing the sensitivity. 

For a binary sequence $A=(b_1,\ldots,b_n)$ and the associated score
sequence $X_A=(x_1,\ldots,x_n)$, 
consider the evolution of the prefix score
$\sum_{i=1}^{k}x_i$ for $k=1..n$. The evolution can be represented as a walk
on $\mathbb{Z}^2$ starting from the origin $(0,0)$ and evolving through
two possible vectors $(1,s)$ and $(1,-p)$. The one-to-one
\begin{figure}[!ht]
\begin{center}
        \includegraphics[width=8cm]{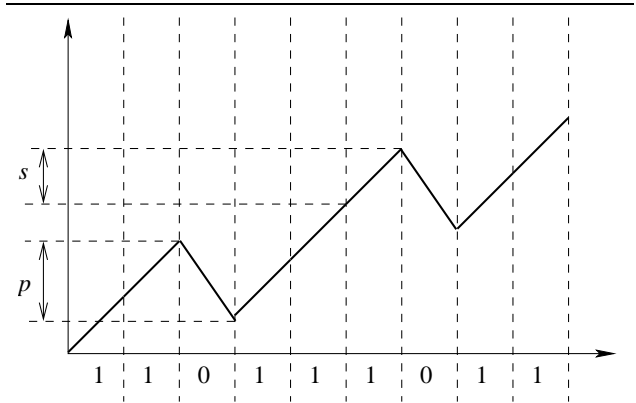}
        \caption{An alignment uniquely associated with a walk. $1$ stands for a match and $0$ for a mismatch}
        \label{bijection}
\end{center}
\end{figure}
relation between a binary sequence and a walk is illustrated in Figure~\ref{bijection}.

Homogeneous sequences correspond to walks $C$ on $\mathbb{Z}^2$ starting at $(0,0)$,
ending at $(n,S)$, and verifying two additional conditions:
\begin{itemize}
  \item $C$ is {\em positive}: $\forall (k,y)\in C, k>0 \Rightarrow y > 0$,
  \item $C$ is \emph{culminating}: $\forall k < n, (k,y)\in C \Rightarrow y < S$.
\end{itemize}
A walk verifying both conditions is called a {\em culminating positive walk}
(CPW). 
It is easily seen that the condition for a walk to be both positive
and culminating is equivalent to the homogeneity of the underlying
sequence. 

We will be interested in two cases, depending on whether the total
score (culminating point) $S$ is fixed or not. We start with the case of sequences of
arbitrary score $S$ and then show how the algorithm is modified to the
case of fixed score. 
Let $C_n$ be the set of all CPWs of size $n$ and arbitrary total score
$S$. 

A classical approach to the random generation of sequences of a given
length drawn from a language $L$ is based on counting suffixes of those
sequences \cite{Wil77}. It allows to generate sequences incrementally from left to
right.
In our case, this
approach is preferable to the generation by rejection. The latter
consists in generating sequences uniformly among all 
possible sequences and discarding those that do not meet the
constraints. 
Although this method also yields a uniform distribution, and
generating each candidate sequence can be done in linear time, the
time complexity of the rejection method heavily depends on $s$ and $p$
parameters. It would be efficient if the constraints are not too strong
-- e.g. when the probability of rejection is $O(1/n)$.
In our setting, if $s$ is smaller than $p$ (which is often the case in practice), the rejection
probability tends to $1$ with an exponential speed as $n$ grows, thus
leading to an expected exponential time complexity of generation.

The counting approach is based on the following general scheme. 
Let $L$ be a set of sequences over an alphabet $\Sigma=\{a_1,\ldots ,a_m\}$
and $L_n$ be the sequences of $L$ of size $n$. 
Let $w_p$ be a prefix of some sequence of $L_n$. We call $P_a(w_p)$ the
probability that $w_p$ can be followed by $a \in \Sigma$ to form a
sequence of $L_n$:
\begin{equation}
P_a(w_p)=\frac{|\{w'|w_p a w' \in L_n\}|}{|\{w''|w_p w'' \in L_n\}|}.
\label{fproba}
\end{equation}
\begin{lemma}
\label{generation}
Given values $P_a(w_p)$ for all $a\in A$ and all prefixes $w_p$, one can
generate sequences of $L_n$ uniformly.
\end{lemma}
\begin{proof}
Starting from $\varepsilon$, issue consecutively letters $\alpha_1,
\ldots,\alpha_n$  with probabilities
  $P_{\alpha_1}(\varepsilon)$, $P_{\alpha_2}(\alpha_1)$, $P_{\alpha_3}(\alpha_1 \alpha_2),
\ldots,P_{\alpha_n}(\alpha_1 \alpha_2 \cdots \alpha_{n-1})$.²
The probability of issuing a sequence $w=\alpha_1 \alpha_2 \alpha_3 \cdots \alpha_n$ is
{\small
  {\setlength\arraycolsep{2pt}
    \begin{eqnarray}\label{f1}\
      &P(\alpha_1 \ldots \alpha_n)  =\nonumber\\
      &P_{\alpha_1}(\varepsilon) P_{\alpha_2}(\alpha_1) P_{\alpha_3}(\alpha_1 \alpha_2)  \ldots P_{\alpha_n}(\alpha_1 \alpha_2 s \alpha_{n\!-\!1})=\nonumber\\
      &\frac{|\{w'|\alpha_1 w' \!\in\! L_n\}|}{|L_n|} \frac{|\{w'|\alpha_1 \alpha_2 w'\!\in\! L_n\}|}{|\{w'|\alpha_1 w' \!\in\! L_n\}|}  \ldots \frac{|\{w'|\alpha_1 \dots \alpha_n w' \!\in\! L_n\}|}{|\{w'|\alpha_1 \cdots \alpha_{n\!-\!1} w' \!\in\! L_n\}|}\!\!=\nonumber\\
      &\frac{1}{|L_n|}
    \end{eqnarray}}}
as $\{w'|\alpha_1 \dots \alpha_n w' \in L_n\} = \{\varepsilon\}$. Therefore, this
yields a uniform generation procedure.  
\end{proof}

\begin{figure}[!ht]
  \begin{center}
    \includegraphics[width=8cm]{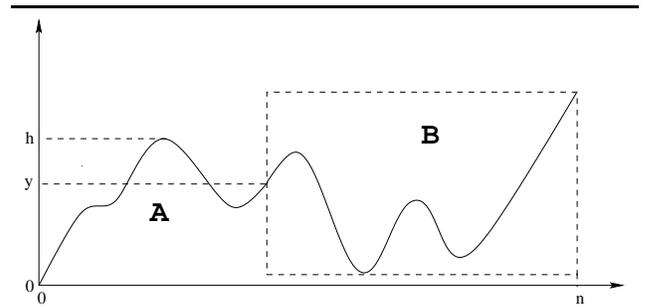}
    \caption{Suffixes \texttt{B} associated to a prefix \texttt{A}
      of $C_n$ depend only of the maximal ordinate $h$ and current ordinate $y$} 
    \label{decomp}
  \end{center}
\end{figure}
In general, one has to precompute up to $|\Sigma|^n$ values $P_\alpha(w_p)$. 
However, in the case of homogeneous sequences, it is not necessary to
process each prefix $w_p$ individually, as 
the only relevant information of $w_p$ is the maximal ordinate $h$ reached by $w_p$ and the current ordinate $y$ (see
Figure~\ref{decomp}).  Therefore, we introduce the concept of 
$(h,y)$-initialized walk.
\begin{definition}
For $h,y\geq 0$, $y\leq h$, an $(h,y)$-initialized CPW is a CPW
starting at $(0,y)$ and culminating
at some point $(n,S)$, such that $S>h$. 
\end{definition}

Let $C_{y,h,n}$ be the set of $(h,y)$-initialized CPWs of length $n$.
\begin{lemma}
\label{nb-suf}
Assume that $w_p$ is a positive walk from $(0,0)$ to $(k,y)$ and $h$
is its maximal ordinate. Then $\{w'|w_p w' \in C_n\} = C_{y,h,n-|w_p|}$. 
\end{lemma}
A proof is immediate and is illustrated on Figure~\ref{decomp}. 

To count the $(h,y)$-initialized CPWs of size $n$ for all
compatible values of $h$,$y$ and $n$, we use the
following recursive decomposition of $(h,y)$-initialized CPWs. A CPW
is represented below as a sequence of vectors $\{(1,s),(1,-p)\}$. 
\begin{lemma}
\label{decomposition}
For $y,h\geq 0$,
\begin{eqnarray}
  C_{y,h,1} =  \left\{ \begin{array}{ll}
          (1,s) & \mbox{if } y+s > h \\ 
          \emptyset & \mbox{otherwise}
          \end{array} \right.
\end{eqnarray}
and for $k>1$, 
\begin{eqnarray}
  C_{y,h,k} = \left\{\begin{array}{ll}
          (1,s)\cdot C_{y+s,\max(h,y+s),k-1} & \\
          ~~\bigcup\,\, (1,-p)\cdot C_{y-p,h,k-1} & \mbox{if  } y > p, \\
          (1,s)\cdot C_{y+s,\max(h,y+s),k-1} & \mbox{otherwise.}\\
          \end{array}\right.
\end{eqnarray}
\end{lemma}
\begin{proof}
A one-step CPW cannot be a $(1,-p)$-step, as it would not be
culminating. It can only be a $(1,s)$-step provided that $h<y+s$. 

The general case is a union of two cases, depending on whether the
first step is $(1,s)$ or $(1,-p)$. The latter is possible only if
$y>p$. If the first step is $(1,s)$, then the new maximum is set to
$\max(h,y+s)$. 
\end{proof}

Note that the decomposition of $C_{y,h,n}$ is unambiguous, which means
that the union operation is a disjoint union. Therefore,
Lemma~\ref{decomposition} gives a recursive formula for computing the
number of CPWs $C_{y,h,k}$ for different values $1\leq k\leq n$,
$0\leq y,h\leq s\cdot n$. A dynamic programming implementation
of the recursion leads to an $O(n^3)$ time and space complexity. 

Lemma~\ref{nb-suf} allows to count the number of possible suffixes for
any prefix $w_p$ of a walk of $C_n$. Let $k=n-|w_p|$, $y$ be the ordinate of the
final point of $w_p$ and $h$ the maximal ordinate reached by $w_p$. 
Then, the probability $P_1(w_p)$ that $w_p$ is followed by a step
$(1,s)$ is $|C_{y+s,\max(h,y+s),k-1}|/|C_{y,h,k}|$. 
As soon as the values $|C_{y,h,k}|$ are computed,
a sequence of $C_n$ is generated incrementally in time $O(n)$ using
Lemma~\ref{generation}. 

\bigskip
We now modify the method to generate homogeneous sequences of 
{\em fixed} score $S$, which amounts to generating CPWs with a fixed
cumulating point. This case is simpler, as there is only a finite
number of possible intermediate scores (states). Therefore the
set of sequences becomes a regular language, for which there exists a
linear-time random generation algorithm (including the preprocessing
time) \cite{HiCo83,Den96b}. In our case, it is sufficient to
precompute an $S\times n$ table 
storing the number of CPW suffixes for each point of the rectangle specified
by corner points
$(0,0)$ and $(n,S)$. Those can be seen as walks inside each rectangle
with corners $(0,y)$ and $(k,S)$. Let $D^{S}_{y,k}$ be the set of such 
walks. The following lemma establishes the corresponding recurrence. 
\begin{lemma}
\label{decomposition2}
For $y\geq 0$,
\begin{eqnarray}
  D^{S}_{y,1} = \left\{ \begin{array}{ll}
          (1,s) & \mbox{if } y+s = S \\ 
          \emptyset & \mbox{otherwise}
          \end{array} \right.
\label{f4}
\end{eqnarray}
and for $k>1$, 
\begin{eqnarray}
  D^{S}_{y,k} = \left\{ \begin{array}{ll}
          (1,s)\cdot D^{S}_{y+s,k-1} & \\
          ~~\bigcup\,\,(1,-p)\cdot D^{S}_{y-p,k-1} & \mbox{if }p\!<\!y\!<\!S\!-\!s,\\
          (1,s)\cdot D^{S}_{y+s,k-1} & \mbox{if } y\!\leq\!p\!<\!S\!-\!s\\
          (1,-p)\cdot D^{S}_{y-p,k-1} & \mbox{if } p<\!S\!-\!s\!\leq\!y\\
          \emptyset & \mbox{otherwise.}
          \end{array} \right.
\label{f5}
\end{eqnarray}
\end{lemma}
Again, the union is disjoint, and therefore the recurrence
can be used for counting the cardinality of each $D^{S}_{y,k}$. Using
Lemmas~\ref{generation} and \ref{nb-suf}, this gives a uniform generation
algorithm of $O(S\cdot n)$ space and time complexity.

\section{Computing the hit probability}
\label{sensitivity}

We present now an algorithm for computing the hit probability on
a random homogeneous sequence, that can be applied to different
seed strategies such as single, double or multiple seeds, contiguous
or spaced seeds, 
etc. 
The algorithm can be seen as an extension of the dynamic programming algorithm of 
\cite{KeichLiMaTromp02} for computing the hit probability for a single seed,
under the Bernoulli model of the sequence. The algorithm of \cite{KeichLiMaTromp02} has
been extended in several ways: \cite{BrejovaBrownVinarCPM03} proposed an extension to
the (Hidden) Markov Models of the sequence; another technique,
proposed in \cite{BuhlerKeichSunRECOMB03}, allows to deal with Markov Models of
the sequence and with multiple seeds; finally, in \cite{BrejovaBrownVinarWABI03}, the
algorithm of \cite{KeichLiMaTromp02} was extended to another seed model, called 
{\em vector seeds}. A similar-style dynamic programming algorithm was proposed in
\cite{BurkhardtKarkkainen03} in a purely combinatorial setting, for computing
the so-called {\em optimal threshold}, which is the minimal number of
seed occurrences for given sequence length and number of substitution
errors (see also \cite{PevznerWaterman95}). 

The extension we propose here is of different nature, as our
probabilistic space is not specified by a probabilistic model of the
alignment, but by a set of possible alignments 
and the condition of the uniform distribution. In other
words, here we impose {\em global constraints} on the alignments (to be
homogeneous and to have a given score) rather than specifying their local properties, as Markov
models do. 
The key
of the construction is the representation from the previous section of
those sequences as random culminating walks on the plane, together with counting
formulas (\ref{f4}), (\ref{f5}). 

For the sake of simplicity of presentation, we describe the algorithm
for a single spaced seed. At the end of this section, we explain how
the algorithm can be extended to multiple-seed strategies. 

Recall that a seed $\pi$ is a string over $\{0,1\}$, where $1$
stands for 'match' and $0$ for a don't care symbol. The length $l$ of $\pi$ is
called the {\em span} of $\pi$ and the number of $1$'s in $\pi$ its
{\em weight}. A seed $\pi$ {\em matches} a string $u\in\{0,1\}$ of length
$l$, if for each position $i$, $\pi[i]=1$ implies $u[i]=1$. A seed $\pi$ 
{\em detects}
a sequence (gapless alignment) $A\in\{0,1\}^*$ if $\pi$ matches some
substring of $A$. 

We now describe a dynamic programming algorithm for computing the
exact probability that a given seed $\pi$ of span $l$, detects a random
homogeneous sequence $A$ of length $n$ and score $S$ under a scoring
system $(s,-p)$. 

  \begin{figure}[htb]\center
    \includegraphics[width = 8.2cm]{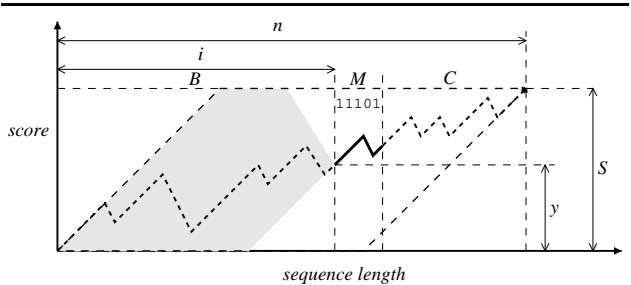}
    \caption[proba]{\label{figure1}Computing the seed detection
      probability on homogeneous sequences}
  \end{figure}
Consider a prefix of a random homogeneous sequence $A$ and assume that
this prefix ends with a suffix $M$. That is, let $A=BMC$ and let $|B|=i$ and $S(B)=y$. 
Let $P(i,M,y)$ be the probability that $\pi$ detects the prefix $BM$ of
a random sequence $A$ (see Figure~\ref{figure1}). Thus, our goal is to
compute the probability $P(n,\varepsilon,S)$ using a set of recursive
equations that we define now. The following are initial conditions
of the recursion:
  \begin{description}
  \item[(i)] $P(i,M,y) = 0$, if $i+|M|<l$,
  \item[(ii)] $P(i,M,y) = 1$, if $|M| = l$ and $\pi$ detects $M$.
  \end{description}
The following conditions insure that the probabilistic space is
respected. Condition (iii) says that the sequences under consideration cannot have a
negative score or a score greater than $S$. Condition (iv) is optional
but allows to cut off at earlier stages some infertile branches of the
computation. It insures that the walks are inside the diagonal band
defined by extremal points (see Figure~\ref{figure1}). 
  \begin{description}
  \item[(iii)] $P(i,M,y) = 0$, if  $y \geq S$ and $i< n$, or  $y \leq 0$ and $n > 0$,
  \item[(iv)] $P(i,M,y) = 0$, if  $y > i\cdot s$ or $y < S -
    (n-i)\cdot s$.
  \end{description}
The following conditions describe main recursion steps. 
  \begin{description}
 \item[(v)] if $\pi$ does not detect $1^{l-|Mb|}Mb$  ($b \in \{0,1\}$),
then $P(i,Mb,y) = P(i,M,y)$,
  \item[(vi)] if $|M| < l$, then $P(i,M,y) = P_1 P(i-1,1.M,y-s)  +
    P_0 P(i-1,0.M,y+p)$, where $P_1$ and $P_0$ are computed using
    formulas (\ref{f4}), (\ref{f5}):
\begin{equation}
\label{probas}
P_1=\frac{|D^S_{S-y+s,i-1}|}{|D^S_{S-y,i}|},\ \ P_0=\frac{|D^S_{S-y-p,i-1}|}{|D^S_{S-y,i}|}
\end{equation}
  \end{description}
Condition (v) says that if $Mb$ is not a suffix of any match of $\pi$,
then the last letter $b$ can be dropped out. 
Condition (vi) is the most tricky one. It says that
if $M$ is shorter than $l$, then the probability is decomposed into
the sum of two terms corresponding to two possible states of the walk
right before the start of $M$ (Figure~\ref{figure1}). A way to compute
the probability $p_0$ and $p_1$ of each of those states is to ``flip over'' the whole
picture and to think of the walk as coming from point $(n,S)$ to
$(0,0)$. Then, $P_0$ and $P_1$ are computed by formula (\ref{fproba})
using the counting technique described in the previous section.
The walks that contribute to probabilities $P_0$ and $P_1$ are those
located inside the shadowed zone in Figure~\ref{figure1}.

The recursive decomposition of $P(n,\varepsilon,S)$ goes as follows: by applying (vi),
the size of $M$ increases up to length $l$, then by alternating (vi)
and (v), the size of $M$ alternates between $l$ and $(l-1)$ while $i$
decreases unless conditions (i)-(iv) apply. 

The worst-case complexity of the algorithm is $O(2^{l}\cdot S\cdot n)$ both
in time and space. The time complexity can be improved by introducing
a preprocessing step and exploiting the structure of the seed $\pi$,
following a general method described in
\cite{KeichLiMaTromp02,BrejovaBrownVinarCPM03}. 
If $w$ is the weight of the seed $\pi$, 
the time complexity can be made $O(l\cdot2^{l-w}\cdot (l^2 + S\cdot
n))$. We refer to \cite{BrejovaBrownVinarCPM03} for details. 

The algorithm presented above can be extended to certain multi-seed
detection
strategies, when a {\em hit} is defined as two or more proximate
occurrences of the seed. A multi-seed strategy is used in Gapped BLAST \cite{GBLAST97}
(two non-overlapping seed occurrences), BLAT \cite{BLAT02} (two or more
non-overlapping occurrences), 
PatternHunter \cite{PATTERNHUNTER02} (two possibly overlapping
occurrences), YASS \cite{NoeKucherovRRINRIA03} (any number of possibly
overlapping occurrences, with additional restriction on the overlap
size). 

To extend the algorithm to the case of $K$ 
seeds without constraints on the overlap, it is sufficient
to perform the recursion on the probability $P(i,M,y,k)$, where the
additional parameter $k$, $0\leq k\leq K$, means that $k$ distinct
occurrences of the seed are assumed to occur in $BM$ (see
Figure~\ref{figure1}). The modification will mainly concern relation
(ii), which will read as $P(i,M,y,k)=P(i,M^-,y,k-1)$, where $M^-$ means
word $M$ without the rightmost letter. 
If the overlap between two successive seeds is upper-bounded by some
constant $\Delta$ (possibly zero, in which case no overlap is
allowed), the modification still holds, except that $M^-$ should be set
to $M$ without $\Delta$ rightmost letters. If the detection strategy
imposes an upper bound on the distance between two neighboring seeds,
the recursion gets more complex, as yet another parameter should be
introduced to ``store'' the distance between the closed seed on the
left. 

\section{Experimental results}
\label{experiments}

To demonstrate the bias introduced by ignoring the property of
homogeneity, we compared the detection probabilities of different
seed strategies on homogeneous vs non-constrained alignments of
a given score and different lengths. The probabilities for homogeneous
alignments were computed by the algorithm of
Section~\ref{sensitivity}. For general alignments, a simpler version
of the algorithm was used, that does not account for the homogeneity
constraint (details are left out). 

\begin{figure*}[htb]\center
  \caption[pr]{\label{pr}Seed detection probability on 
    homogeneous vs arbitrary alignments}
  \includegraphics[width = 8cm]{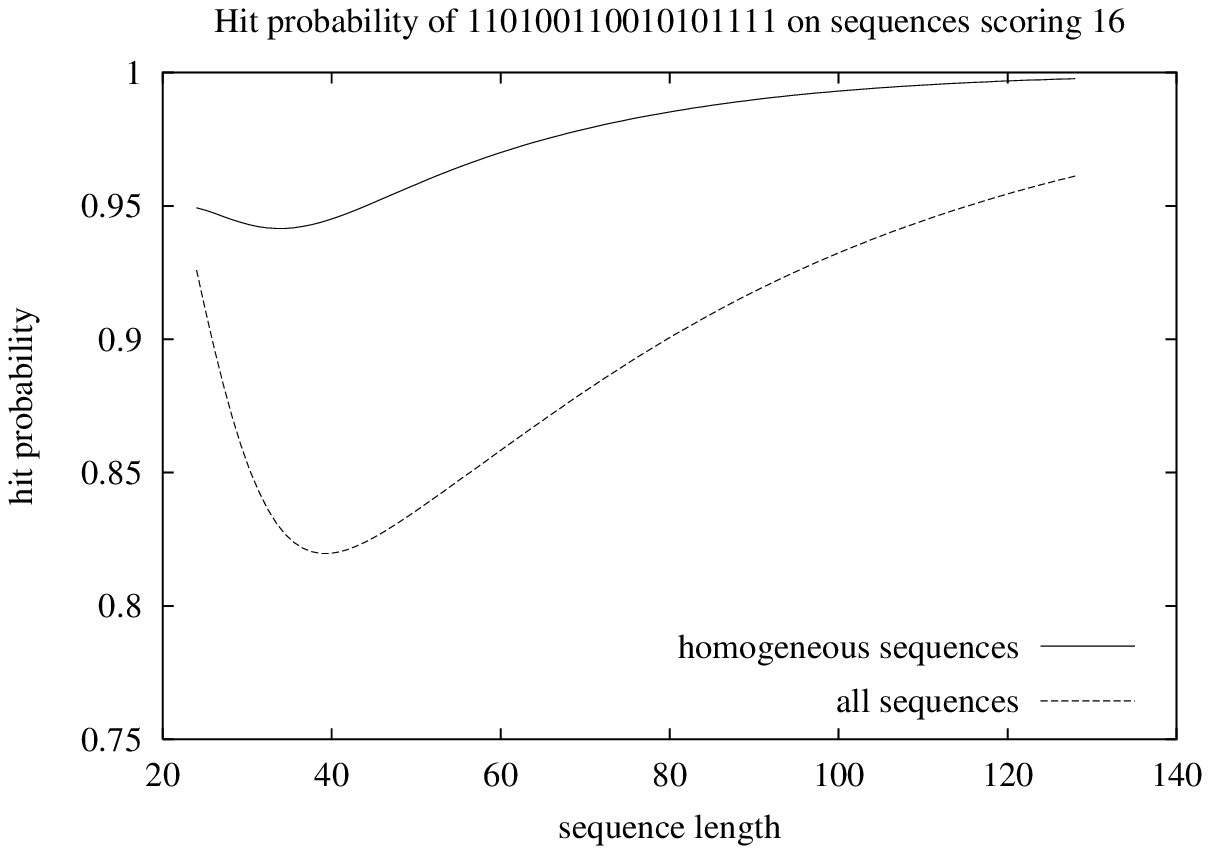}\includegraphics[width = 8cm]{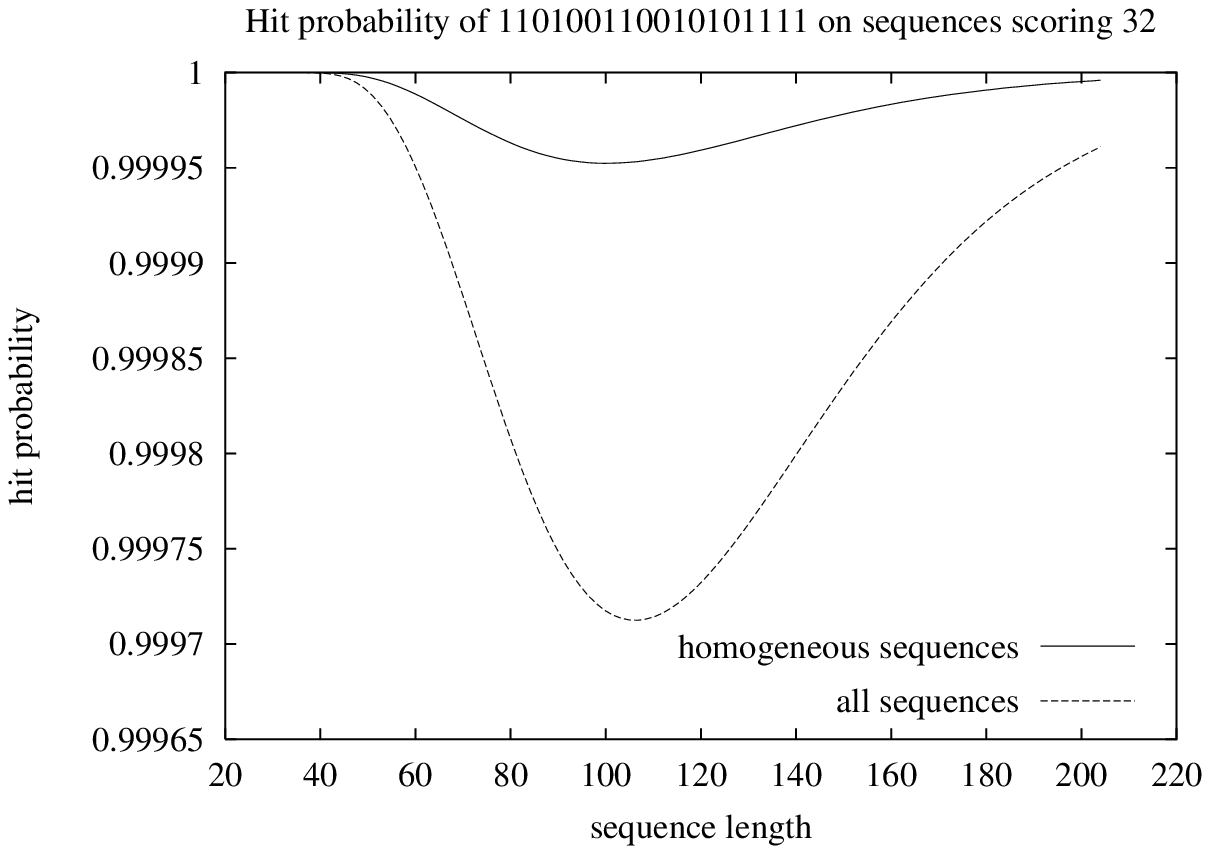}
  \includegraphics[width = 8cm]{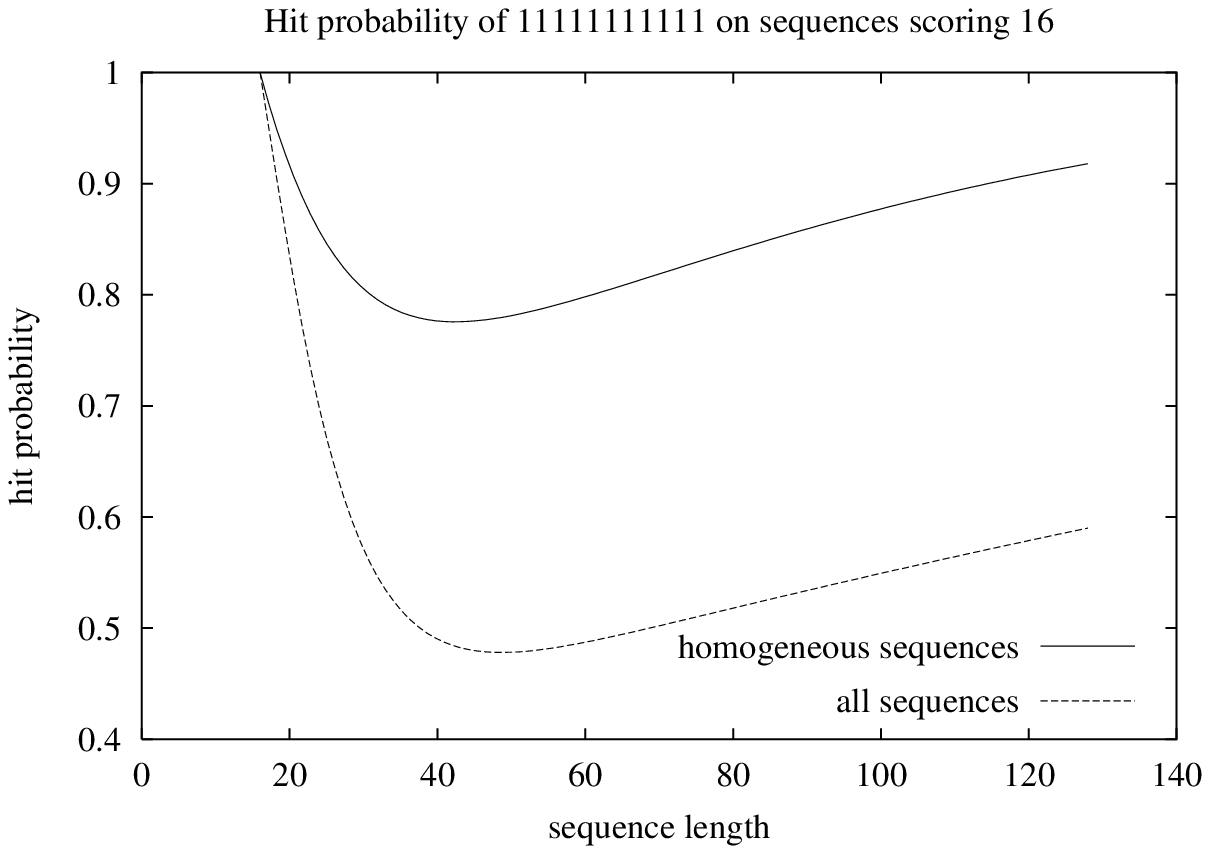}\includegraphics[width = 8cm]{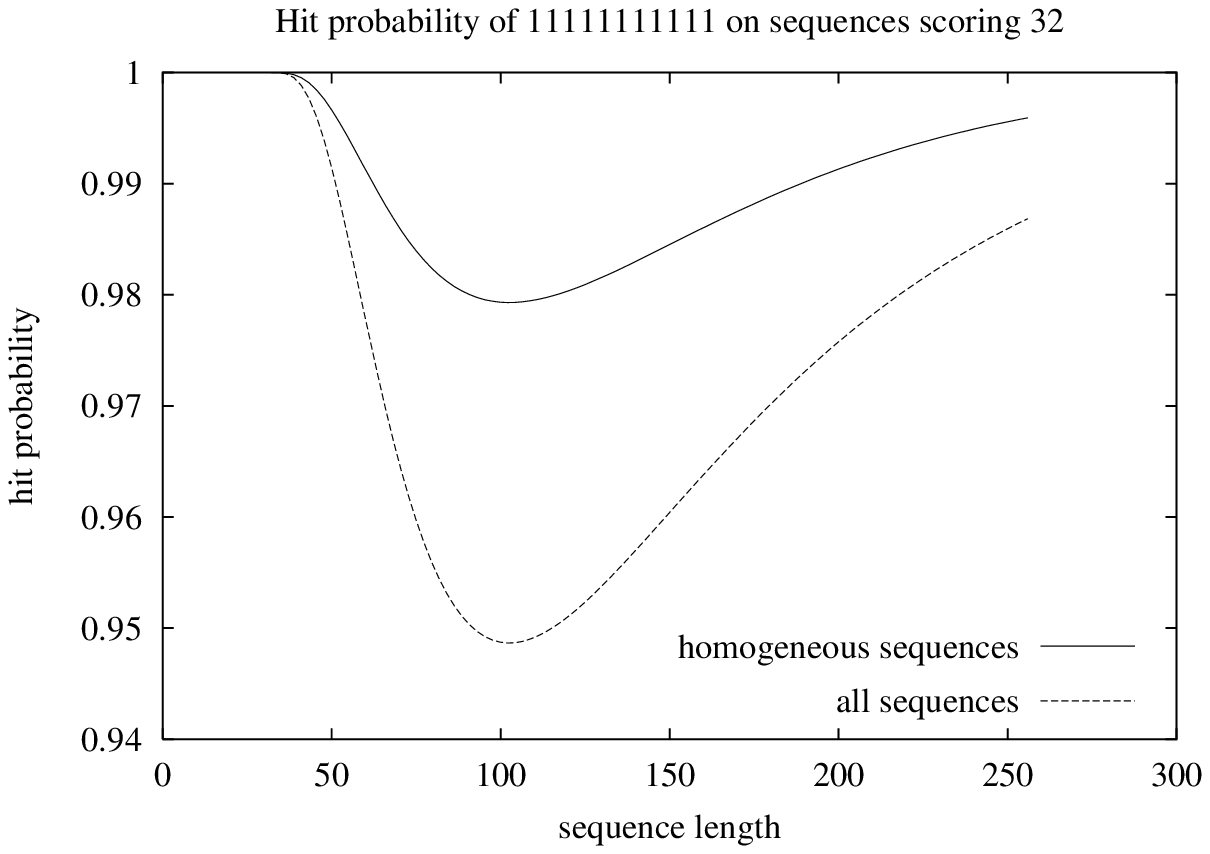}
\end{figure*}
Figure~\ref{pr} shows the results. Each plot represents the
probability of a certain seed to detect homogeneous/arbitrary
alignments of a certain score as a function of their length. 
All experiments reported in this section use the default
BLAST +1/-3 scoring system. Left and right plots correspond to score
16 and 32 respectively. The upper row corresponds to the seed
110100110010101111 of weight 11 and span 18 (implemented
in PatternHunter \cite{PATTERNHUNTER02}), while the lower row
corresponds to the contiguous seed of weight 11 (implemented in
BLAST 1 \cite{BLAST90}). 

For all settings, the results clearly show that ignoring the condition
of homogeneity leads to a considerable underestimation of the
sensitivity. The fraction of homogeneous alignments missed is, in most
cases, at least two times less than what one would expect out of
measurements on arbitrary alignments. 

One of the most important applications of measuring the sensitivity is
the design of optimal spaced seeds for the detection of alignments of a
given type. Therefore, we made another group of experiments aiming
at comparing the most sensitive seeds for homogeneous vs arbitrary
alignments. Some results are shown in Table~\ref{seeds}. 

We computed {\em optimal seeds} for detecting homogeneous and arbitrary alignments of length
$40$, for several score values (between $12$ and $24$). Some results are shown in Table~\ref{seeds}. They have been
obtained by an exhaustive search through all seeds of span up to 20.
The probabilities were computed by the algorithm of
Section~\ref{sensitivity}. The table shows
that for the same parameters (alignment score and seed weight), 
the optimal seeds are different, depending on whether the
optimality is defined with respect to all alignments (probability
$P_a$) or only homogeneous ones (probability $P_h$). In each case, the
highest probability is shown in slanted characters.

\def\SHAPE#1{\mbox{~\scriptsize{{\tt #1}~}}}
\begin{table}[htb] %
  \center%
  {\small%
    \begin{tabular}{ll|lll}
      $(n,S)$ & $w$ & $\pi^{\{h|a\}}_{w}{\scriptstyle(n,S)}$ & $\mathcal{P}_h $ & $\mathcal{P}_a$  \\
      \hline \hline
      $(40,12)$ & 9  & \SHAPE{1110010110111}    & {\textsl{0.986271}} &          0.902372   \\ 
      $(40,12)$ & 9  & \SHAPE{111001001010111}  &          0.983516   & {\textsl{0.917869}} \\
      \hline
      $(40,16)$ & 9  & \SHAPE{1110010110111}    & {\textsl{0.998399}} &          0.988887   \\
      $(40,16)$ & 9  & \SHAPE{1100110101111}    &          0.998353   & {\textsl{0.989535}} \\
      \hline      
      $(40,16)$ & 10 & \SHAPE{11101100101111}   & {\textsl{0.98742}}   &          0.938499   \\
      $(40,16)$ & 10 & \SHAPE{110110010101111}  &          0.98740     & {\textsl{0.942769}} \\
      \hline
      $(40,20)$ & 10 & \SHAPE{11101001110111}   & {\textsl{0.999172}} &          0.996303   \\
      $(40,20)$ & 10 & \SHAPE{110110010101111}  &          0.999065   & {\textsl{0.996555}} \\
      \hline                 
      $(40,20)$ & 11 & \SHAPE{111011101001111}  & {\textsl{0.975462}} & {\textsl{0.993076}} \\
      \hline                 
      $(40,24)$ & 11 & \SHAPE{111010011110111}  & {\textsl{0.999891}} & {\textsl{0.999661}} \\ 
    \end{tabular}}%
  \caption{\label{seeds} Optimal seeds for 
    homogeneous vs arbitrary alignments}
\end{table}%


Furthermore, we have performed a similarity search based on seeds from
Table~\ref{seeds}, using YASS
software~\cite{NoeKucherovRRINRIA03}. In particular, we compared the number of alignments
found using the seed optimized on homogeneous alignments vs the one
for arbitrary alignments. Table~\ref{exp} shows the results for the
seeds from Table~\ref{seeds} of weight 9 and 10
($\pi^{h}_{w}(n,S)$, respectively $\pi^{a}_{w}(n,S)$, stands for the
optimal seed from Table~\ref{seeds} computed on
homogeneous and all alignments respectively). The experiments were
made on comparing full chromosomes IV (1560kb), V (580kb), IX (450kb), XVI (960kb)  of 
{\em Saccharomyces Cerevisiae} against each other or against
themselves. Both strands of each chromosome has been processed in each
experiment ({\tt -r 2 } option of YASS). The search was done with the
``group size'' parameter 10 and 11 for seed weight respectively 9 and 10 (option
{\tt -s} of YASS). 
The results show that, in most cases, the seed tuned for homogeneous
alignments allows to identify more relevant similarities than the seed
optimized for all alignments.

\def\SHAPE#1{\mbox{\scriptsize{\tt #1}~~}}
\begin{table}[htb] %
  \center%
{\small%
    \begin{tabular}{r|rr|rr}
& \multicolumn{2}{c|}{weight $9$} & \multicolumn{2}{c}{weight $10$} \\
      sequences &
      $\pi^{h}_{9}{\scriptstyle(40,12)}$ & $\pi^{a}_{9}{\scriptstyle(40,12)}$ &
      $\pi^{h}_{10}{\scriptstyle(40,20)}$ & $\pi^{a}_{10}{\scriptstyle(40,20)}$ \\
      \hline \hline
      IX       &  519  &  519  &  502  &  496 \\
      IX / V   &  364  &  356  &  342  &  329 \\
      IX / XVI &  408  &  387  &  383  &  348 \\
      IX / IV  &  523  &  521  &  488  &  473 \\
      V        &  500  &  487  &  477  &  466 \\
      V  / XVI &  961  &  955  &  937  &  891 \\
      V  / IV  & 1273  & 1258  & 1248  & 1192 \\
      XVI      &  539 &  554 &  545  &  510 \\
      XVI / IV & 1429 & 1448 & 1452  & 1368 \\
      IV       & 1542  & 1539  & 1510  & 1461 \\
      \hline
       \end{tabular}}%
  \caption{\label{exp} Number of high-scoring alignments found with a
    seed optimized for homogeneous alignments (left column) vs that
    optimized for all alignments (right column)}
\end{table}%

\section{Discussion}
\label{conclusions}

In this paper we presented an approach to measure the sensitivity of
seed-based similarity search strategies. The main
point is to compute the hit probability over homogeneous alignments,
rather than arbitrary alignments. The property of homogeneity
requires that the alignment does not contain significant
negative-score segments occurring either inside the alignment (in
which case the alignment can be decomposed into alignments of higher
score), or at the edges (in which case a subalignment should be
considered). 
In this paper, we showed that ignoring this property leads to a bias
in estimating the detection capacity of seeds. 

Recently proposed approaches to estimate the seed detection probability
\cite{BuhlerKeichSunRECOMB03,BrejovaBrownVinarCPM03,BrejovaBrownVinarWABI03}
assume a Markov model of alignment, that specifies its 
{\em local} composition. 
The approach of this
paper is complementary, as we only impose global constraints
(homogeneity, total score) and abstract from local properties of
the alignment. If we want to account for local properties, 
the assumption that all fixed-score homogeneous
sequences are equiprobable would be no longer justified.

Note that 
one of the drawbacks of the Markov model approach of
\cite{BuhlerKeichSunRECOMB03,BrejovaBrownVinarCPM03,BrejovaBrownVinarWABI03}
is that the alignment score is taken into account only
indirectly, through the expected composition of the alignment
(or {\em identity rate}, in case of the match/mismatch model). 
This is a serious
disadvantage if one has to measure the probability on alignments of
different length (as in \cite{NoeKucherovRRINRIA03}), since in this case
the same score generally corresponds to different identity rates. 
The approach proposed in this paper is based on the score rather than
on the identity rate.

Our analysis has been based on the match/mismatch model of
alignment. However, sometimes it is very useful to distinguish between
different mismatches, 
for example between transitions and transversions
(see the footnote 
in Section~\ref{enumeration}). This approach leads to
modeling alignments by sequences on non-binary alphabets (e.g. a
three-letter alphabet of match/transition/transversion). 
From a pure computer science point of view, the results of
Sections~\ref{enumeration},\ref{sensitivity} can be extended to the
non-binary alphabet. However, from the biological point of view,
the assumption of the uniform distribution over all
sequences becomes unrealistic in this setting, as letters are
obviously no more ``equivalent'', and properties of alignment
composition must be added to the model. 
To conclude, the ultimate approach should take into account both global
properties of the alignment and its local and compositional
properties. Designing such an approach remains a challenging problem. 

\paragraph{Acknowledgments} We are grateful to Alain Denise for his
help. We also thank Mikhail Roytberg for comments and helpful
discussions. 
G.~Kucherov and L.~No{\'e} have been supported by the french
{\em Actions Sp{\'e}cifiques  ``Algorithmes et S{\'e}quences'' and
  ``Indexation de texte et d{\'e}couverte de motifs''}  of CNRS. Yann
Ponty has been supported by the french 
IMPG group {\em ``Algorithmique, combinatoire et statistiques des s{\'e}quences
g{\'e}nomiques''}

\bibliographystyle{ieee}
\bibliography{paper}

\end{document}